\documentclass[utf8,11pt]{article}

\usepackage{url,hyperref,lineno,microtype,subcaption,amssymb,graphicx}
\usepackage[onehalfspacing]{setspace}

\def\Authors{Juste Raimbault\,$^{1,2,3,\ast}$}
\def\Address{$^{1}$ CASA, University College London\\
$^{2}$UPS CNRS 3611 ISC-PIF\\
$^{3}$UMR CNRS 8504 G{\'e}ographie-cit{\'e}s}

\def\corrEmail{$^{\ast}$juste.raimbault@polytechnique.edu}

\begin{document}

\title{\vspace{-3cm}An agent-based model of interdisciplinary interactions in science} 

\author{
\Authors\medskip\\
\Address\medskip\\
\corrEmail
} 

\date{}

\maketitle

\begin{abstract}

An increased interdisciplinarity in science projects has been highlighted as crucial to tackle complex real-world challenges, but also as beneficial for the development of disciplines themselves. This paper introduces a parcimonious agent-based model of interdisciplinary relationships in collective entreprises of knowledge discovery, to investigate the impact of scientist-level decisions and preferences on global interdisciplinarity patterns. Under the assumption of simple rules for individual researcher project management, such as trade-offs between invested time overhead and knowledge benefit, model simulations show that individual choices influence the distribution of compromise points between emergent level of disciplinary depth and interdisciplinarity in a non-linear way. Different structures for collaboration networks may also yield various outcomes in terms of global interdisciplinarity. We conclude that independently of the research field, the organization of research, and more particularly the local balancing between vertical and horizontal research, already influences the final positioning of research results and the extent of the knowledge front. This suggests direct applications to research policies with a bottom-up leverage on the interactions between disciplines.

\end{abstract}

\section{Introduction}

The role of interdisciplinary projects in science has been highlighted as crucial for the development of complexity approaches and an effective tackling of real-world issues. Many aspects of knowledge production have a role in enhancing interdisciplinary collaborations. \cite{Hofstra9284} study the circular relationship between diversity and innovation, and show that underrepresented groups have a higher likelihood of successfully innovate in science. \cite{jang2019coevolutionary} use an agent-based model to study the co-evolution between knowledge diffusion and the structure of knowledge. Each discipline has its own view on interdisciplinarity, as for example \cite{urbanska2019does} unveil an asymmetry between social and hard sciences in the credit given to other disciplines within interdisciplinary projects. Other social or political factor are to be taken into account when investigating the disciplinary structure of science: access to funding has for example a strong impact on the efficiency of knowledge production \cite{gross2019contest}. \cite{akerlof2018persistence} show that the discrepancy between disciplines is intrinsic to the type of knowledge produced, as they suggest that paradigms are more likely to persist in ``low-power'' sciences. The organisation of research is also an important factor, and teams and single authors produce different aspects of the common knowledge \cite{pavlidis2014together}. \cite{rouse2018modeling} model probables trajectories according to the type of research environment. The link between open access, which is a driver of increased collaborations and potentially increased interdisciplinarity, and the quality of research, is investigated by \cite{VANVLOKHOVEN2019751}.

Interdisciplinarity in itself has extensively been studied by quantitative studies of science. \cite{2019arXiv191003628T} show that interdisciplinary papers perform better in terms of citation on the long run than mainstream papers. \cite{zeng2019increasing} investigate the interdisciplinarity of scientists themselves and how it evolved in time, and show that more scientists have switched between topics recently. \cite{lariviere2010relationship} provide empirical evidence for an optimal intermediate level of interdisciplinarity in terms of research impact.
\cite{brown_murray_furlong_coco_dablander_2020} study within the particular context of an interdisciplinary summer school the propensity of mixing within interdisciplinary projects, and find evidence consistent with random mixing. \cite{pluchino2019exploring} show that randomness has an important role in determining individual trajectories success in physics.


Following \cite{giere2010agent}, agent-based modeling is a privileged approach to simulate the behavior of scientists. \cite{shafiee2019agent} use an agent-based model to simulate the impact of a workflow to process data under different collaboration scenarios. \cite{doi10.1162qssa00008} simulate citation dynamics, and more particularly the consequence of introducing a performance index on citation patterns. Agent-based modeling has extensively been used for the evaluation of peer review practices. \cite{feliciani2019scoping} surveys 46 simulation studies of peer review with numerous applications. \cite{kovanis2016complex} empirically calibrates an agent-based model of peer review for more than 100 journals, and provides a tool to evaluate systems of peer reviews. \cite{shneiderman2018twin} describes a theoretical model involving various actors of science. Agent-based models are more broadly used to study social dynamics such as group organisation in \cite{dionne2019diversity}.


Various works have dealt with microscopic modeling of knowledge production, among which for example the Nobel game introduced by~\cite{chavalarias2016s} which investigates the balance between falsification of previous theories and the elaboration of new theories. \cite{giere2010agent} also proposed an agent-based model of science, consistently with the perspectivist approach developed in \cite{giere2010scientific}. We develop here a simple agent-based model of scientific research focusing on the interplay between disciplinary and interdisciplinary research. The rationale relies on the basic assumption that scientists can choose when starting a new project between interdisciplinary collaboration and a work within their discipline. How can the choice patterns at the micro-level influence the overall interdisciplinarity level ? The model is voluntary parcimonious to test if even many simplification some structural effects still hold.

\section{An agent-based Model of Interdisciplinarity}

\subsection{Rationale}

Many dimensions and processes are at play to shape collaborations between scientists and more broadly between scientific disciplines. These include for example social networks, governance and funding issues, or knowledge proximity (which can occur on various knowledge domains, from methodological to empirical or theoretical). Our rationale is to propose an agent-based model grasping some of this complexity from the bottom-up focusing on scientist behavior, but simple enough so that it can be systematically explored. We include thus in the model two basic antagonist processes, namely a propensity to collaborate mostly determined by knowledge proximity, and some resources constraints (time, funding) which affect negatively the possibility to collaborate. Working with scientists outside one's field has indeed a high cost, from finding common ground and research questions to an possible construction of integrated knowledge \cite{frodeman2013sustainable}.

\subsection{Model description}

Agents are $N$ scientists $A_i$, characterized by a probability distribution $d(x)$ representing their disciplinary positioning in an abstract way: research is summarized by a one dimensional variable $\mathbb{R}$, and the disciplinary positioning on this axis is given by the distribution. The model is setup with normal distributions of width $\sigma$ with an average distributed uniformly in $\left[0;1\right]$. Scientists also have a time budget per day, that we will summarize as a future timetable $T(t_0):t>t_0 \mapsto p(t) \in \mathcal{P}$ where $\mathcal{P}$ is the space of scientific projects. The central feature of the model is the utility function $U(d_i,d_j)$ determining an abstract utility for scientist $i$ to collaborate with $j$ for a given project. It will be a function of the disciplinary overlap $o = \int_x d_i(x)\cdot d_j(x) dx$ and different assumptions on the form of this cost function can be tested. We take a linear cost in the overlap and a varying benefit, expressing the fact that researchers have different strategies regarding their interdisciplinary positioning. This way, we have $U(d_i,d_j) = o / i^\alpha - o$, assuming a fat-tail distribution of individual preferences for interdisciplinarity, given by a power law of parameter $\alpha$. A discrete choice formulation gives the probabilities for a scientist $i$ to choose among $j$ collaborators by $p_j = \exp\left(\beta U(d_i,d_j) \right)/\sum_k \exp\left(\beta U(d_i,d_k) \right)$. Given a social network of relations, that we take for now as a fixed scale-free social network, the temporal evolution of the model goes as follows: (i) one scientist with no current activity is picked up at random, and starts a project with one of its potential collaborators taken as its neighbors in the network that have free time, chosen with the probability $p_j$. The project has a random uniform duration and timetables are updated accordingly; (ii) current projects are updated and finished if necessary. The outcome of the model if measured by average depth across project, defined for one project as the overlapping areas between distribution, and average interdisciplinarity measured by total area covered.

\section{Results}

\subsection{Empirical data}


\begin{figure*}
\includegraphics[width=0.49\textwidth]{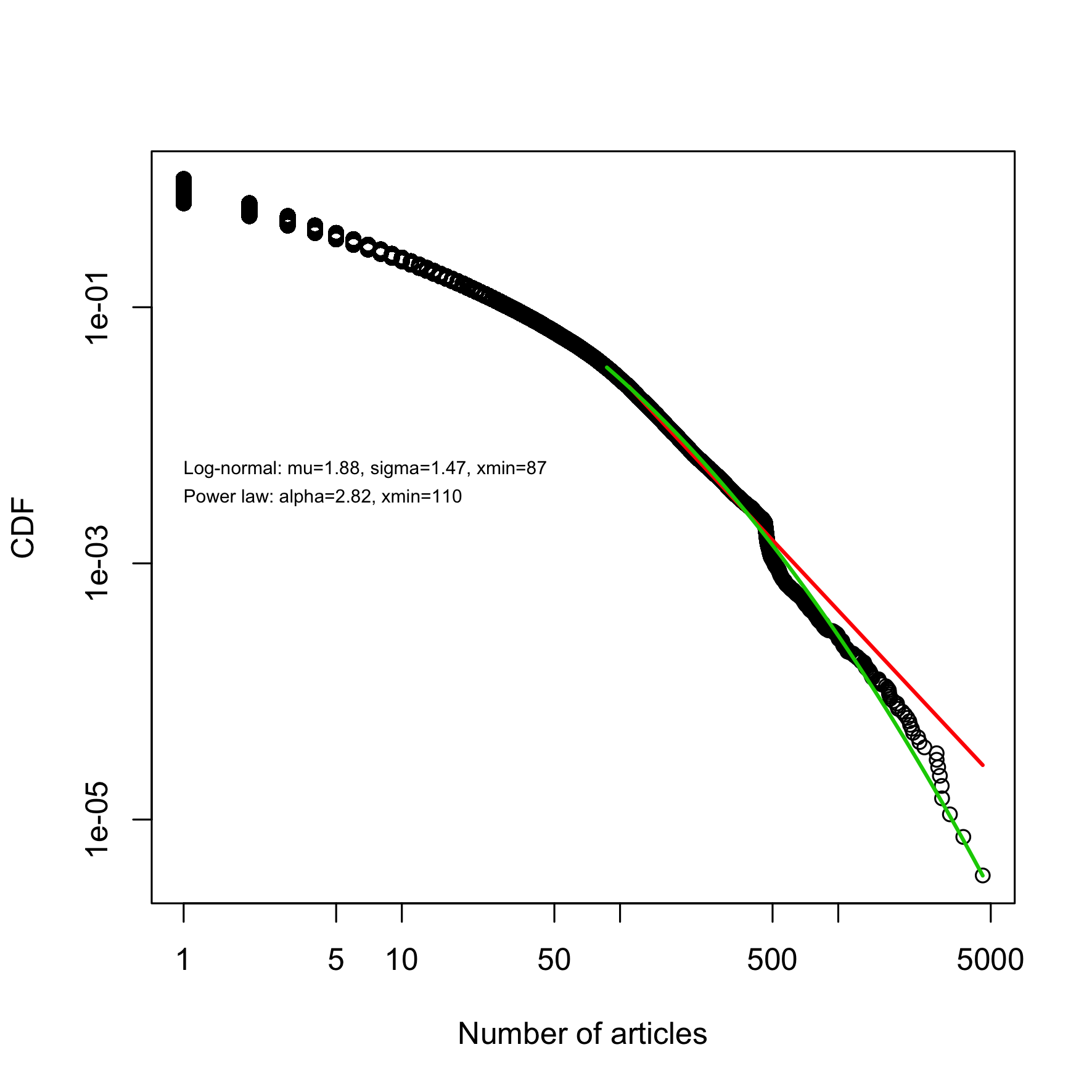}
\includegraphics[width=0.49\textwidth]{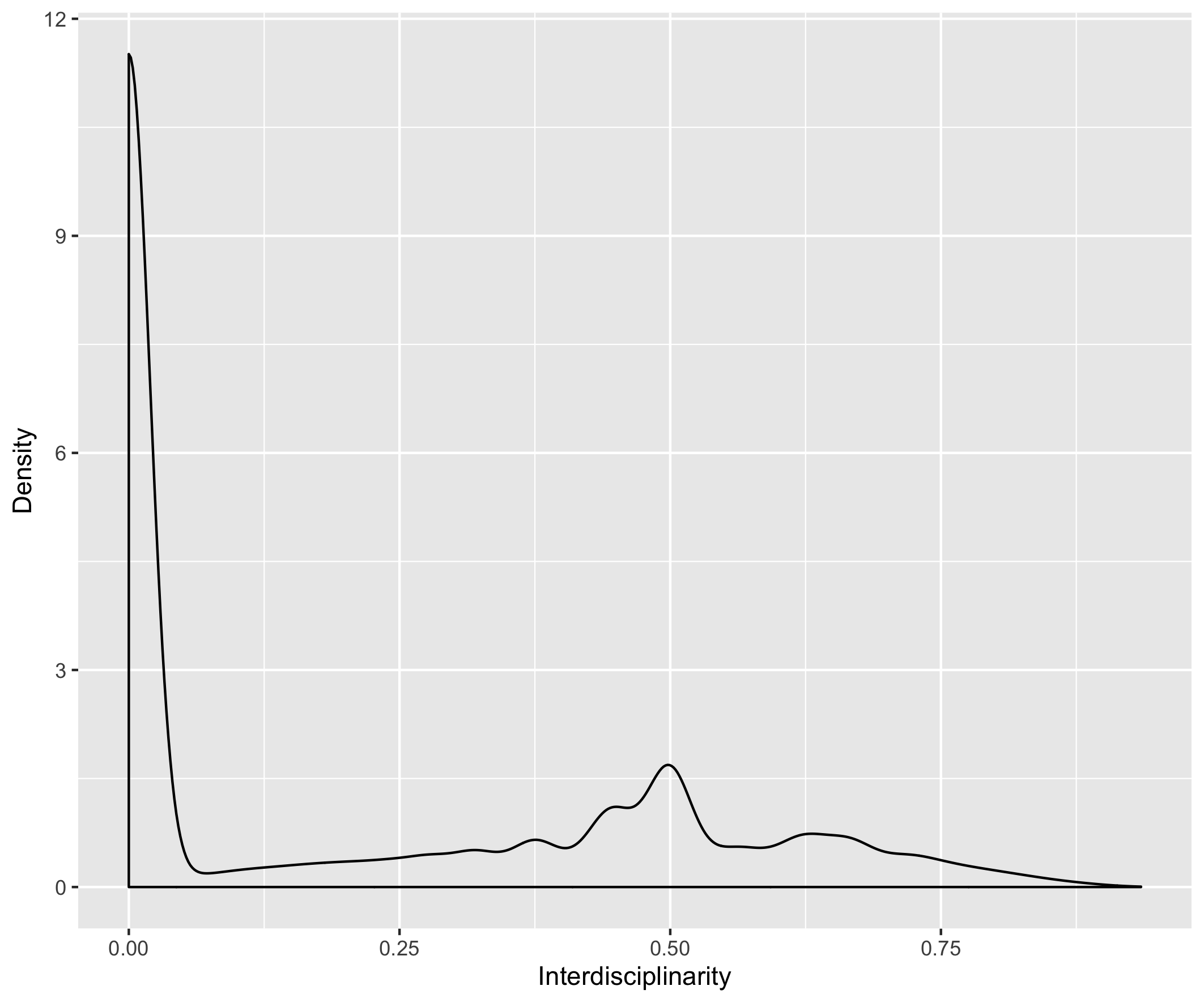}\\
\includegraphics[width=0.49\textwidth]{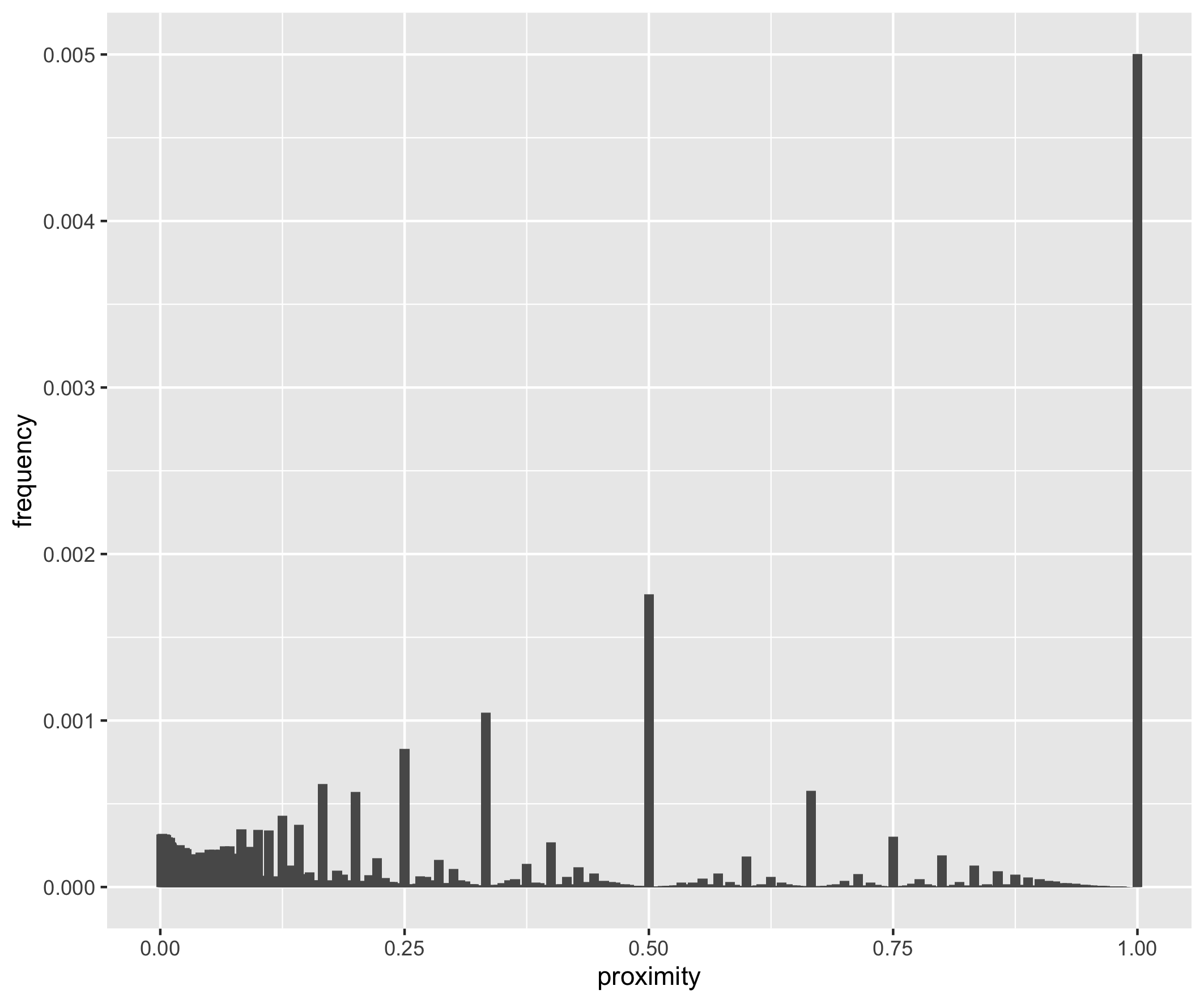}
\includegraphics[width=0.49\textwidth]{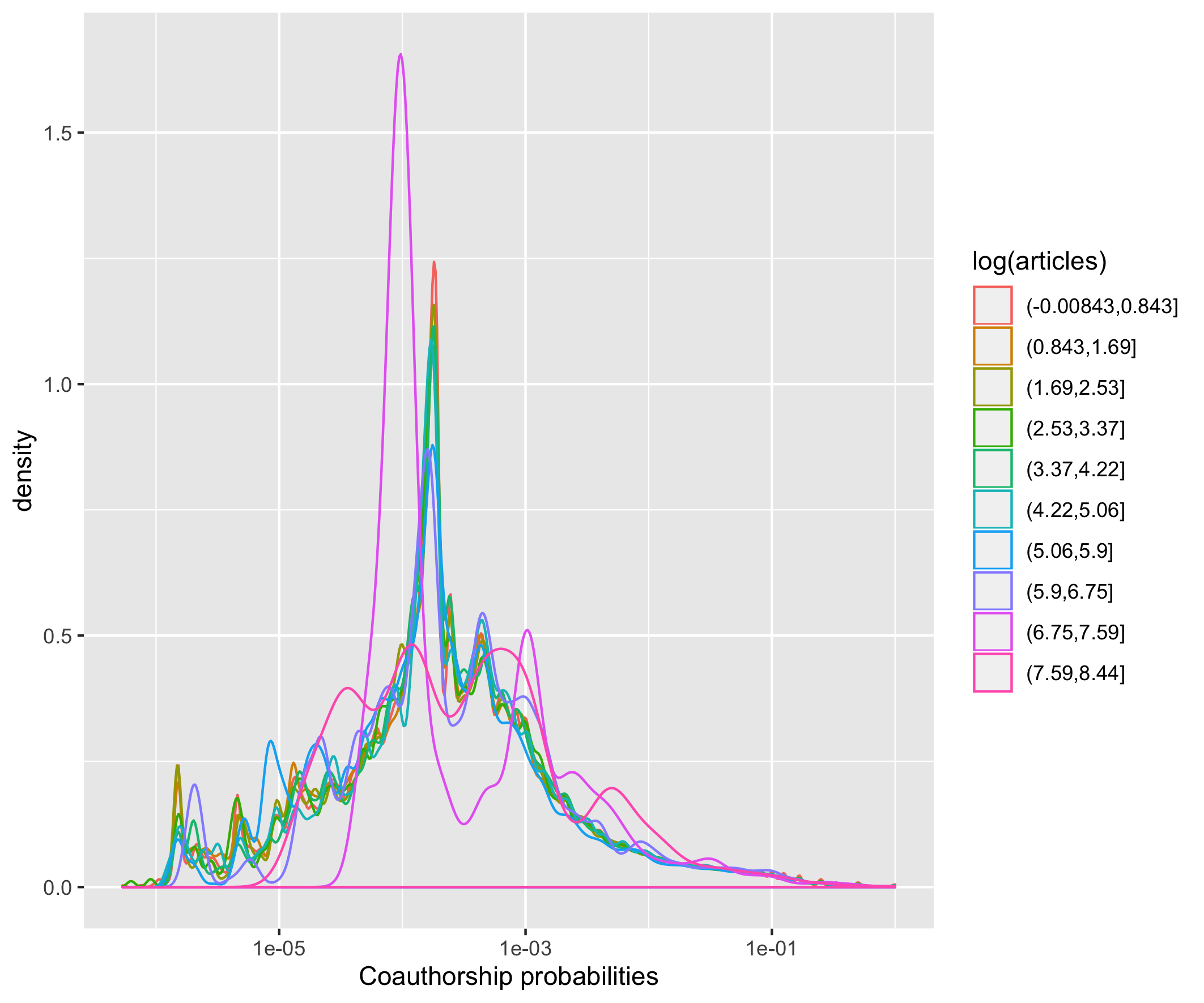}
\caption{\textbf{Collaborations and interdisciplinarity within the Arxiv dataset.} \textit{(Top left)} Cumulative distribution function of the number of articles per author (these were disambiguated using first and last name only, statistics may not be accurate). We compare a log-normal and a power-law fit. \textit{(Top right)} Distribution of interdisciplinarity per author, computed as an Herfindhal index of probabilities within endogenous citation communities. \textit{(Bottom left)} Distribution of positive author proximities, defined as cosine similarity between authors probability distribution within citation communities. \textit{(Bottom right)} Distribution of co-authorship probabilities, conditioned by the number of articles.\label{fig:empirical}}
\end{figure*}

In order to give empirical support to the modeling choices for the ABM, we first study the properties of a large scientific corpus. We propose to use the Arxiv citation network, which represents a significant proportion of physics and computer science. An open dataset providing parsed authors and citations is made available by \cite{clement2019use}. This allows constructing a citation network with $\left|V\right| = 1,396,261$ nodes (papers) and $\left|E\right| = 6,849,633$ citation links. This corresponds to $1,506,500$ unique authors which we disambiguated by concatenating first name and last name. We then proceed to a community detection in the citation network, using a Louvain community detection algorithm. We obtain therein a modularity of $0.78$ and 38 communities with a size larger than 1000. Working with these main endogenous citation communities (which can be interpreted as scientific fields of citation practice), we construct probabilities for authors to belong to each community. These are computed as $p_{ik} = N_{ik} / N_{i}$ for author $i$ and community $k$, were $N_{ik}$ is the number of articles authored within this community and $N_i$ the total number of articles authored. This allows computing a cosine proximity between authors defined as $s_{ij} = \vec{p}_i \cdot \vec{p}_j$, and also an interdisciplinarity measure as an Herfindhal diversity index given by $h_i = 1 = \sum_k p_{ik}^2$. Finally, we also study co-authorship probabilities $c_{i \rightarrow j}$ defined as the probability for author $i$ to co-author with author $j$ knowing that the author has written a paper (the matrix is thus non symmetric).

We show in Figure~\ref{fig:empirical} the empirical results obtained. The number of papers by author is close to a power-law with an exponent of 2.82, although a log-normal law seems to better fit the data. Regarding interdisciplinarity of authors, although a large majority of authors are mono-disciplinary, we find a secondary peak at 0.5 and a non negligible proportion of authors spanning the indicator range up to very high values of 0.8. This confirms the relevance of our model with an active interdisciplinarity. When studying cosine similarity between authors using their probabilistic description within communities, we find a broad range of values, also witnessing a high diversity (knowing that most authors are at a 0 proximity, since the plot is conditional for readability). Co-authorship probabilities follow rather symmetrical distributions with fat tails on a log-scale, consistently when conditioning on the number of papers authored. This is consistent with the power-law assumed for the propensity for interdisciplinarity for authors.

\subsection{Model exploration}

The model is implemented in NetLogo \cite{tisue2004netlogo} and explored with OpenMole \cite{reuillon2013openmole}. Source code and results are available on the open git repository of the project at \url{https://github.com/JusteRaimbault/Perspectivism}. Data used in the paper is available on the dataverse at \url{https://doi.org/10.7910/DVN/GMQ5A8}.

\begin{figure*}
	\centering
	\includegraphics[width=0.49\linewidth]{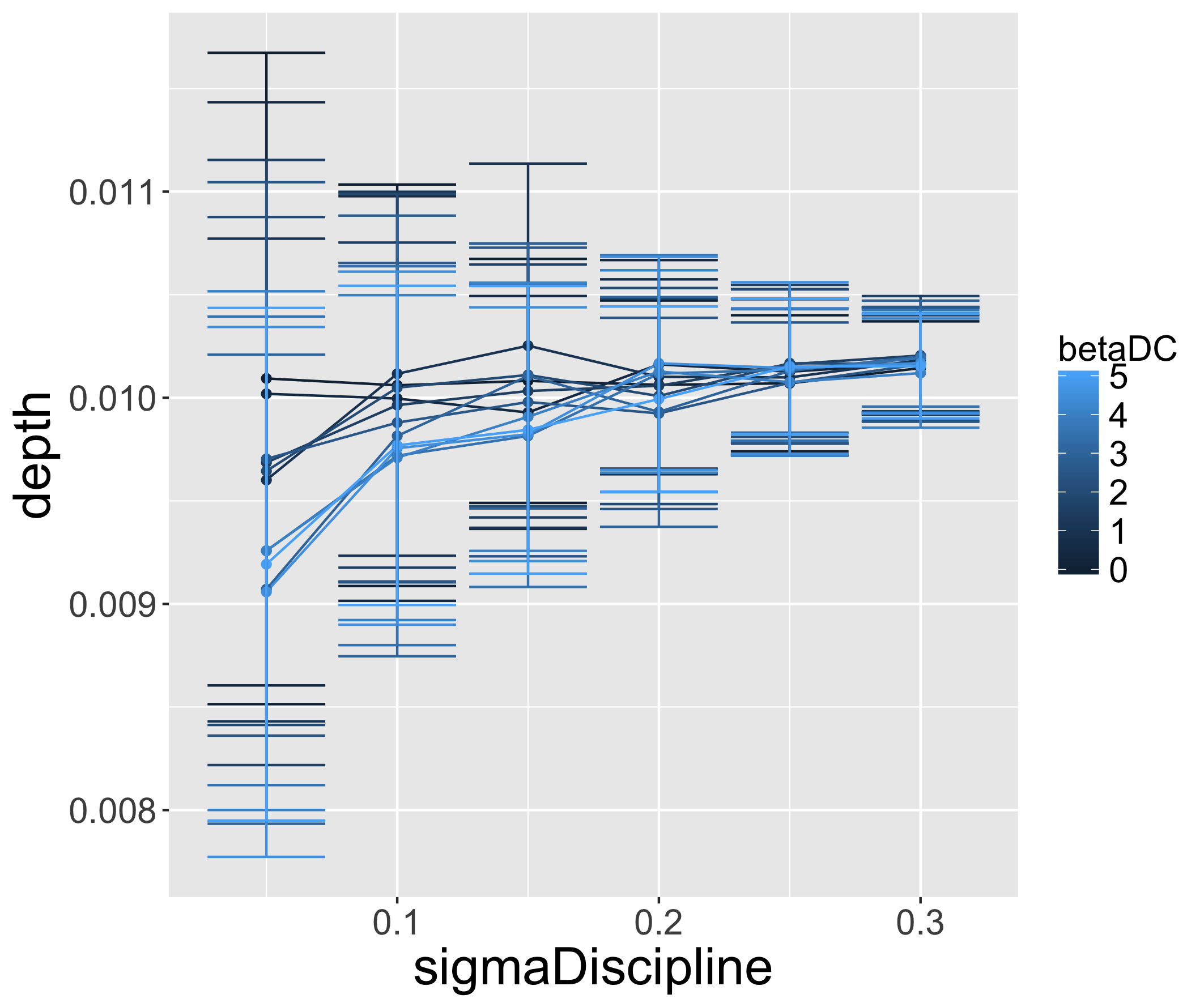}
	\includegraphics[width=0.49\linewidth]{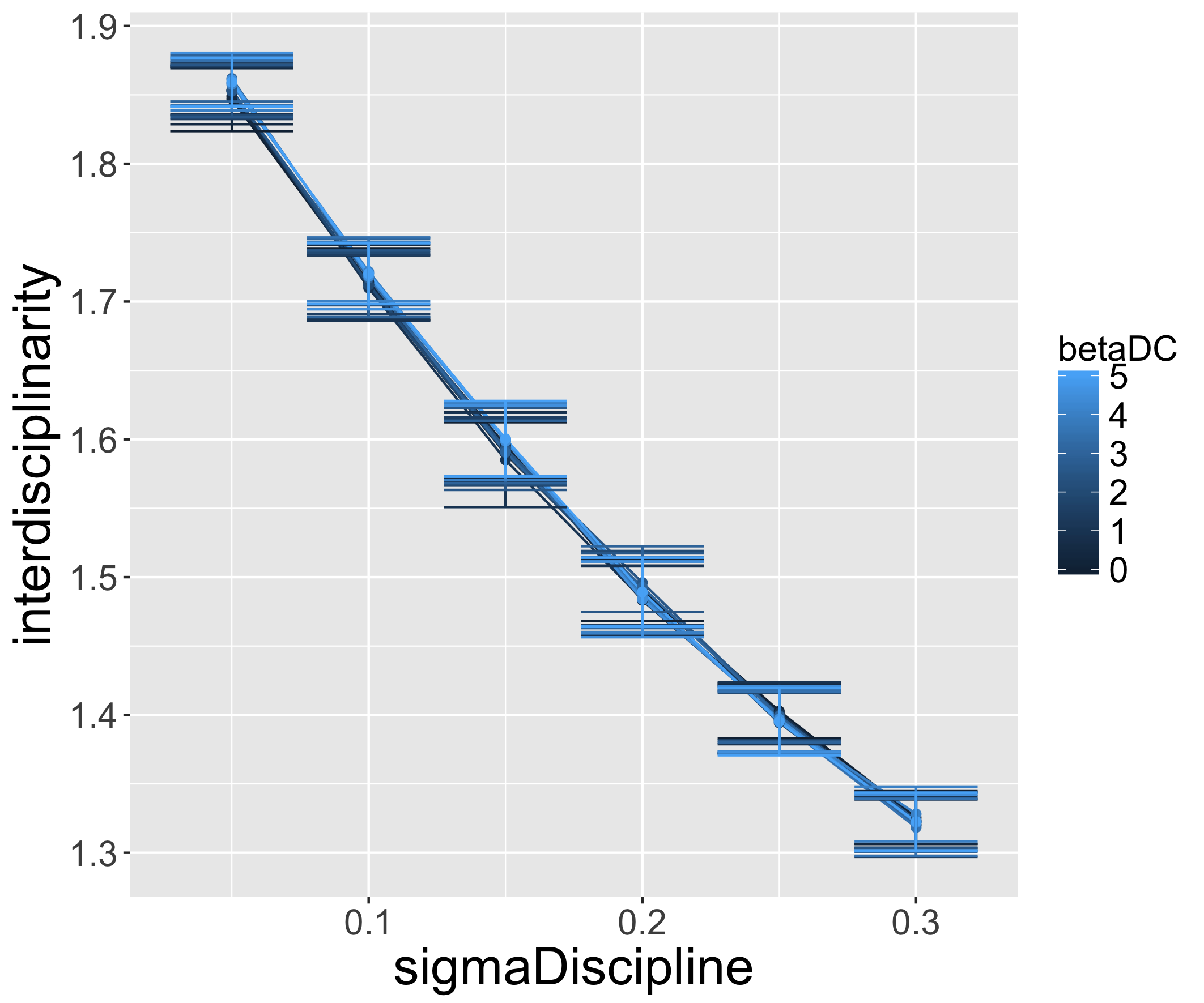}\\
	\includegraphics[width=0.8\linewidth]{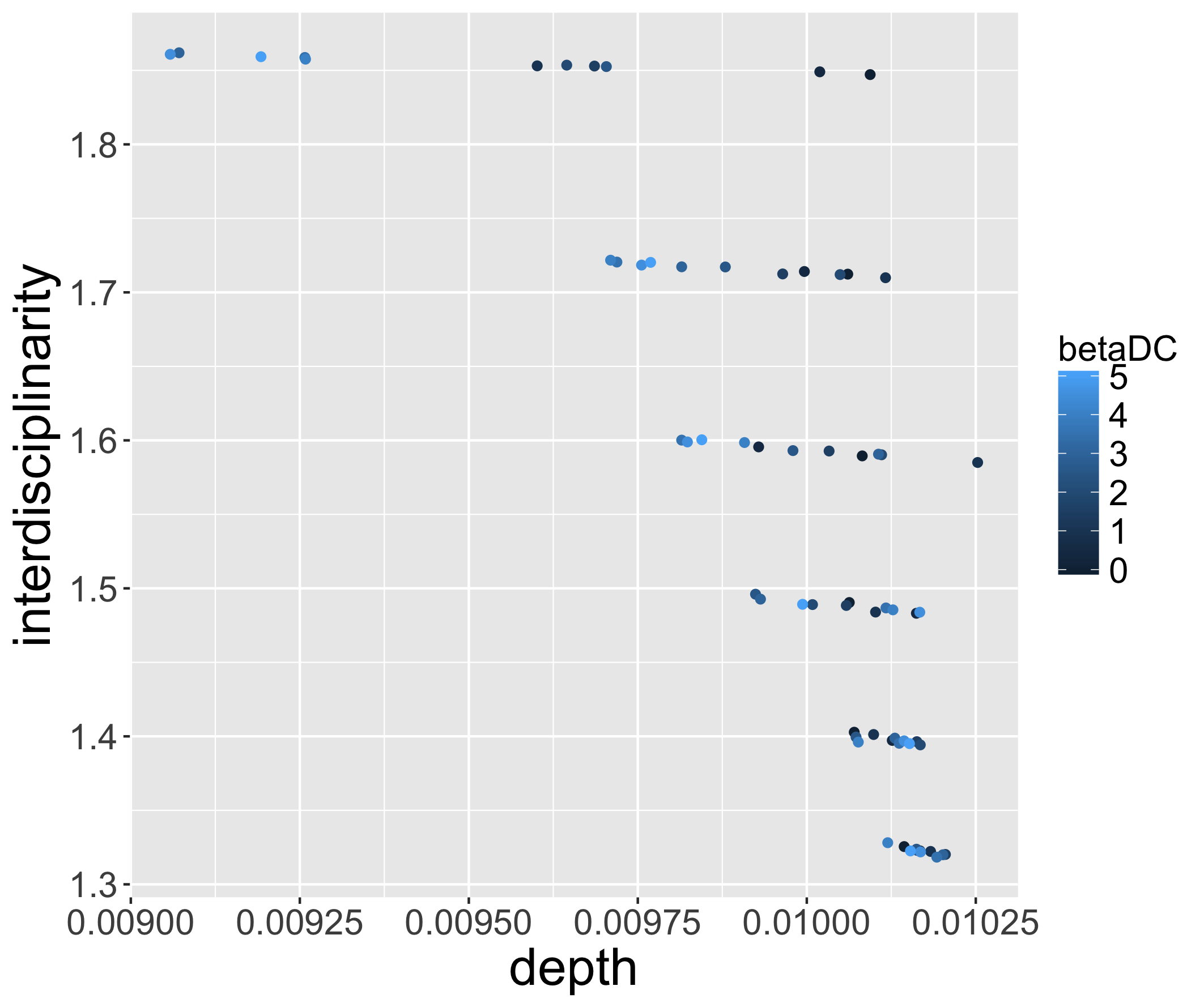}
	\caption{\textbf{Patterns of interdisciplinarity from model simulations.} We show measures of depth and interdisciplinarity (top row) at fixed $\alpha=0.5$ and network structure, for varying discrete choice parameter $\beta$ as a function of individual extent $\sigma$. On the bottom, the Pareto front of average point between these two objectives.}
	\label{fig:plots}
\end{figure*}

We run a basic grid exploration of the parameter space, both with random and small-world social networks, for parameters $\alpha,\beta,\sigma$ with 50 repetitions of the model for each parameter points, corresponding to 158,400 model runs. Figure~\ref{fig:plots} shows indicators variation on a given subspace and the corresponding Pareto front between depth and interdisciplinarity. We show a second order influence of preference hierarchy $\alpha$ and non-linearity of model behavior as a function of all parameters. Convergence properties are reasonable with this number of repetitions. Large individual disciplinary width $\sigma$ causes the choice parameter $\beta$ to have no influence, whereas low values give an increasing interdisciplinarity and a decreasing depth as a function of $\beta$. Random behavior ($\beta = 0$) leads to a constant depth of projects. When examining the Pareto front between the two contrary objectives, the optimal points occur for intermediate $\beta$ when $\sigma$ is fixed, suggesting non-trivial behavioral optima at a fixed disciplinary configuration. These first exploration show the complex dynamics of interdisciplinarity even with simple interaction rules and network structure, and suggests further applications such as the exploration of policies by changing network structure or studying in a more refined way the influence of $\alpha$. Preliminary non-systematic model experiments, in particular changing the type of network structure, suggest that it may also have significant effect on model outcomes.

\section{Discussion}

\subsection{Perspectivism and Model Coupling}

Beyond the simplifying opposition between fully constructivist and realistic approaches to science, several alternatives have been developed, among which Perspectivism \cite{giere2010scientific} is a way to tackle most of the issues opposing these two by taking an agent-based approach to the production of scientific knowledge. The main feature of this viewpoint is to consider each scientific enterprise as a single perspective, in which an agent aims at understanding an aspect of the real world (the ontology) with the mean of a medium, which is considered as a model. Constituted disciplines thus contains more or less compatible perspectives. The explicitation of this approach has been done by~\cite{raimbault2017knowledge} to embed it into knowledge domains, as a generalization of knowledge domains introduced by~\cite{livet2010ontology}.

We postulate that this approach to science may be a powerful tool to foster interdisciplinary collaborations, if used in a reflexive way in the construction of projects. \cite{Ellemers7561} propose a similar framework. More precisely, we suggest to apply an ``Applied Perspectivism'', in the sense of an explicit perpectivist positioning within a given collaboration, and associated guidelines and protocols for collaboration. This would imply a high-level of reflexivity for each agent implied, a mapping of the different layers of the enterprise and the positioning of each agent regarding the domains of knowledge. This way, in the particular case of model coupling, the explicitation of positioning and of the structure of each knowledge implied should ease interactions. As Banos points out~\cite{banos2013pour}, transversal work must alternate with deeper investigations in each discipline, in a kind of ``virtuous circle'' \cite{banos2017knowledge}. Fostering a synergy between complementary knowledge is the core aspect more important than interdisciplinarity in itself \cite{leydesdorff2020measurement}. This raises the issue of, before individual researcher particularities, how a given collective structure of scientific knowledge production should balance between these disciplinary and interdisciplinary knowledge. It is clear that this question is deeply endogenous to each studied subject, and even each particular approach taken, but within the applied knowledge framework described above, we have reasons to believe that certain structural properties may be rather general. Indeed, each discipline is expected to bring components for each knowledge domain, and the co-evolving perspective is built on their interrelations. This paper proposed to investigate basic aspects of this issue, by means of agent-based modeling.

This work aimed at providing quantitative evidence of the feasibility of the epistemological point of view described above and inform potential implementation for some of its processes, more precisely how can certain level of coupling of perspectives (or overlap of ontologies) may be achieved given specializations of scientists and a given dynamic of interaction.

\subsection{Possible extensions}

Possible refinements of the model, towards a less stylized and more behavioral and micro-based model, could for example include the introduction of time budgets, simultaneous projects and dynamical time investment for scientists. The assumption of two-person projects is also strongly constraining, and relaxing it would require the extension of depth and interdisciplinarity measures that is not necessary straightforward. Furthermore, the absence of learning and of evolution of the social network when completing a project suggests a short time scale of application: further refinements should include dynamics of individual distributions and of individual relationships.

\section{Conclusion}

In conclusion, we show with a simple model that the individual choices produce an emerging structure of the research front, suggesting that applied perspectivism requires a careful tuning of research structure and researcher behaviors since Pareto-optimal configurations correspond to non-trivial parameter points. Future developments should include more realistic behavioral assumption, and a formalisation of the applied perspectivism approach to include it in the agent-based model.

\end{document}